\documentclass[preprint,12pt]{elsarticle}

 \usepackage{graphicx}

\usepackage{amssymb}
\usepackage{amsmath}
\numberwithin{equation}{section}
\usepackage{amsthm}

\journal{Computers \& Mathematics with Applications}

\begin{document}

\begin{frontmatter}



\title{Waves and  distributions connected to 
systems of interacting populations}


\author[a,b]{Nikolay K. Vitanov}
\ead{vitanov@imbm.bas.bg}
\author[c]{Zlatinka I. Dimitrova}
\author[a]{Kaloyan N. Vitanov}

\address[a]{Institute of Mechanics, Bulgarian Academy of Sciences,
Acad. G. Bonchev Str., Bl. 4, 1113 Sofia, Bulgaria}
\address[b]{Max-Planck Institute for the Physics of Complex Systems, Noetnitzerstr. 38,
01187 Dresden, Germany}
\address[c]{"G. Nadjakov" Institute of Solid State Physics, Bulgarian Academy of Sciences
Blvd. Tzarigradsko Chaussee 72, 1784 Sofia, Bulgaria}

\begin{abstract}
We discuss two cases that can be connected to the dynamics of interacting populations: 
(I.)  density waves  for the case or negligible random fluctuations of the populations densities,
and (II.) probability distributions connected to the model equations for  
of spatially averaged populations densities for the case of
significant random fluctuations of the independent quantity that can be associated with the population density. 
For the case (I.) we consider model 
equations containing polynomial nonlinearities. Such nonlinearities can arise as a consequence of 
interaction among the populations (for the case of large population densities) or as a result of a 
Taylor series expansion (for the case of small density of interacting populations). In the both cases
we can apply the modified method of simplest equation to obtain exact traveling-wave solutions connected to 
migration of population members. Such solutions are obtained for systems consisting 
of 1 or 3 populations respectively. For the case (II.)  we discuss model equations of the 
Fokker-Planck kind for the evolution of the statistical distributions of population densities. 
We derive several stationary distributions for the population density and calculate the expected exit 
time connected to the extinction of the studied population. 
\end{abstract}

\begin{keyword}
 interacting populations \sep nonlinear waves \sep method of simplest equation 
\sep density fluctiations \sep probability density functions for populations densities


\end{keyword}

\end{frontmatter}


\section{Introduction}\label{s1}
Since the famous paper of the properties of the Lorenz attractor half a century ago \cite{lorenz}
the nonlinearities are intensively studied in different areas of science. Just several
examples are \cite{kang} from the optics, \cite{siegel} from biomechanics, \cite{campbell}
from solid state physics, \cite{chomaz}-\cite{v98} from fluid mechanics, \cite{schreiber,ks}
from the time series analysis, etc. Population dynamics is a classic area of application of
nonlinear models \cite{turchin,var}.
In many cases the dynamics of interacting populations is studied on the basis of mathematical
models consisting of equations that contain only time dependence of the population
densities \cite{murr}-\cite{dv5}. These models are very useful for understanding 
the complex dynamics of the interacting populations but they do not account for two important 
aspects of this dynamics:
(I.) the possible influence of spatial characteristics of the environment; and (II.)
the possible fluctuations of the population densities caused by different factors.
Below we shall investigate two kinds of population dynamics models that account for
each of these effects. First of all we shall discuss the dynamics of spatially distributed
populations and this will be a continuation of our previous work \cite{vit09,vit09a}.
Then we shall show that by appropriate averaging the spatial model can be
reduced to model in which the population densities depend only on the time. The models
that contain spatially averaged quantities are valid for the cases of small and large values 
of the densities of the interacting populations. Because of this  in the case of the
model for the evolution of the  spatially averaged population densities we shall discuss the influence 
of random fluctuations of the population densities
for arbitrary values of the densities. 
The result of the action of these fluctuations is that instead of equations for the
trajectories of the populations in the phase space of the population densities we shall 
write and solve equations for the probability density functions of the population 
densities.
\par
The organization of the paper is as follows. In Sect. 2 we obtain the model equations for 
a system of interacting spatially distributed populations. In Sect. 3 after appropriate
averaging the model system of nonlinear PD Es from Sect. 2 is reduced to a system of
nonlinear ODEs for the temporal evolution of the densities of interacting populations.
In Sect. 4 we obtain traveling-wave solutions of the system of model equations that can
be associated with density population waves. In sect. 5 we discuss the
influence of fluctuations on the evolution of population densities of the model system of
spatially averaged equations from Sect. 3. The inclusion of random fluctuations transforms the
model system of deterministic nonlinear ODEs to a system of Langevin equations which is
further transformed to a system of Fokker-Planck equations for the evolution of the
probability density functions (p.d.f.s)  of interacting populations. We discuss several
examples for stationary p.d.f.s that are attractors for the time dependent p.d.f.s for the 
case of large times. Several expected exit times connected to extinction of populations are
calculated too. In addition to the main text of the paper there are 3 appendices that are
devoted to: (I) the modified method of simplest equation used to obtain the traveling-wave 
solutions; (II) description of a MAPLE program for obtaining the analytic form of the
solution for the coupled kink waves for the case of 3 interacting populations; and (III) 
remarks on two observations connected to the diffusion Markov processes and the 
Fokker-Planck equation.
\section{Model equations}
\subsection{Spatially distributed populations}
Let us consider an two-dimensional area $S$ where $N$ competing populations are present.
The density of each population is $\rho_i (x,y,t) = \frac{\Delta N_i}{\Delta S}$, where
$\Delta N_i$ is the number of the individuals of the $i$-th population that are present
in the small area $\Delta S$ at the moment $t$. Now let a movement of population members
through the borders of the area $\Delta S$ be possible and let $\vec{j}_i (x,y,t)$ be the
current of this movement. Then $(\vec{j}_i \cdot \vec{n}) \delta l$ is the net number of 
members from $i$-th population, crossing a small border line $\delta l$ with normal vector
$n$. Let the density changes (other then these caused by the border crossings) be summarized
by the function $C_i(\rho_1,\rho_2,\dots,\rho_N,x,y,t)$. Then the change of the density 
of members of the $i$-th population in the studied area is described by the equation
\begin{equation}\label{eq1}
\frac{\partial \rho_i}{\partial t} + {\rm div} \vec{j}_i = C_i .
\end{equation}
Below we shall discuss the case where $\vec{j}_i$ has the form of linear multicomponent
diffusion. In this case
\begin{equation}\label{eq2}
\vec{j}_i = - \sum_{k=1}^N D_{ik} (\rho_i, \rho_k,x,y,t) \nabla \rho_k ,
\end{equation}
where $D_{ik}$ is the diffusion coefficient. Eq. (\ref{eq2}) reflects the possibility
that the motion of the population members is caused not only by gradients of the
density of the own population but also it could be caused by gradients of the densities
of the other populations. 
\par
We shall not specify the
kind of the function $C_i$ as we shall consider the general case of relatively
small populations densities that allow us to write $C_i$ as Taylor series
expansion around the zero values of all population densities as follows
\begin{eqnarray}\label{eq3}
C_i (\rho_1,\rho_2,\dots,\rho_N)= \sum_{n_1=0}^\infty \sum_{n_2=0}^\infty \dots
\sum_{n_N=0}^\infty \alpha^{(i)}_{n_1,n_2,\dots,n_N}
\rho_1^{n_1} \rho_{2}^{n_2} \dots \rho_{N}^{n_N} ,
\end{eqnarray}
where the constant coefficients $\alpha^{(i)}_{n_1,n_2,\dots,n_N}$ are as follows
\begin{eqnarray}\label{eq4}
\alpha^{(i)}_{n_1,n_2,\dots,n_N} = \frac{1}{n_1 ! n_2 ! \dots n_N !}  
\frac{\partial C_i^{n_1 + n_2 + \dots + n_N}}{\partial
\rho_1^{n_1} \partial \rho_2^{n_2} \dots \partial \rho_N^{n_N}}   \Bigg \vert_{\rho_1 = \rho_2 =\dots = \rho_{N}=0} .
\end{eqnarray}
We shall discuss the one-dimensional case and in addition we shall assume that the diffusion
coefficients $D_{ik}$ are constants. Then the substitution of Eqs. (\ref{eq2}) and (\ref{eq3})
in (\ref{eq1}) leads to the following system of nonlinear PDEs for the studied $N$ 
interacting populations:
\begin{eqnarray}\label{eq5}
\frac{\partial \rho_i}{\partial t} - \sum_{k=1}^N D_{ik}  \frac{\partial^2 \rho_k}{\partial x^2} =
\sum_{n_1=0}^\infty \sum_{n_2=0}^\infty \dots
\sum_{n_N=0}^\infty \alpha^{(i)}_{n_1,n_2,\dots,n_N} 
\rho_1^{n_1} \rho_{2}^{n_2} \dots \rho_{N}^{n_N} .
\end{eqnarray}
For the case of 1 population the system (\ref{eq5}) is reduced to the equation
\begin{equation}\label{eq5a}
\frac{\partial \rho}{\partial t} -  D  \frac{\partial^2 \rho}{\partial x^2} =
\sum_{n_1=0}^\infty  \alpha_{n_1} \rho^{n_1} .
\end{equation}
\par
Let us give two examples of systems of kind (\ref{eq5}). The first example is connected
to the classical Lotka-Volterra case extended in \cite{dv2} - \cite{dv4}.
In this case after assumption of dependence of the growth rates and competition coefficients
on the population density one arrives at the system of equations \cite{dv2}-\cite{dv4}
\begin{eqnarray}\label{eq5b}
\frac{\partial \rho_i}{\partial t} - \sum_{j=1}^n D_{ij} \Delta \rho_j = r_i^0 \rho_i
\bigg[ 1 - \sum_{j=1}^n (\alpha_{ij}^0 - r_{ij} \rho_j) - \sum_{j,k=1}^n \alpha_{ij}^0 (\alpha_{ijk} + \rho_k)
\rho_j \rho_k - \nonumber \\
\sum_{j,k,l=1}^n \alpha_{ij}^0 r_{ik} \alpha_{ijl} r_{ik} \alpha_{ijl} \rho_j \rho_k \rho_l  \bigg]
\end{eqnarray}
where $i=1,2,\dots,n$  is a number that indexes the $n$ competing populations; $D_{ij}$ is the
diffusion coefficient; $\Delta = \frac{\partial^2}{\partial x^2} + \frac{\partial^2}{\partial y^2}$;
$r_i^0$ and $\alpha_{ij}^0$ are the parts of the corresponding growth rates and competition
coefficients that do not depend on the population densities; $r_{ik}$ and $\alpha_{ijk}$ are
parameters that regulate the intensity of the dependence of the population growth rates and
competition coefficients on the population densities $\rho_i$. It can be easily checked that the
system described by Eq.(\ref{eq5b}) is a particular case of the system of equations  given by
Eqs.(\ref{eq1})-(\ref{eq4}).
\par
A system of kind similar to the system from Eq.(\ref{eq5b}) arises in the social dynamics
in the spatial model of ideological struggle developed in \cite{va}. For this case the model system
of equations is
\begin{eqnarray}\label{eq5c}
\frac{\partial \rho_i}{\partial t} - \sum_{j=1}^n D_{ij} \Delta \rho_j = r_i \rho_i + \sum_{j=0}^n f_{ij} \rho_j +
\sum_{j=0}^n \alpha_{ij} \rho_i \rho_j + \sum_{j,k=0}^n b_{ijk} \rho_i \rho_j \rho_k + \dots
\end{eqnarray}
where $\Delta = \frac{\partial^2}{\partial x^2} + \frac{\partial^2}{\partial y^2}$; $\rho_i, i=1,2,\dots,n$
are the spatial densities of the populations of the followers of the corresponding ideology; 
$r_i$ are the rates of change of corresponding populations of adepts by births and deaths;
$f_{ij}$ is the coefficient of non-contact conversion (the ideology of a person can be changed
without contact between humans but by the mass media influence for an example); $a_{ij}$ is
the coefficient of binary contact conversion that describes the change of ideology by contacts
between followers of different ideologies.
\subsection{Spatially averaged equations}
Below we shall apply spatial averaging to the system of equations (\ref{eq5}) similar
to the averaging used in the optimum theory of turbulence \cite{ott1} - \cite{ott3}.
In the general two-dimensional case
let a quantity $q(x,y,t)$ be defined in an large two-dimensional plane area $S$ with acreage
$\mid S \mid$.
Then by definition the spatial average of $q$ is
\begin{equation}\label{eq6}
\overline{q}(t) = \frac{1}{\mid S \mid} \int \int_{S} dx \ dy \ q(x,y,t)  .
\end{equation}
$q(x,y,t)$ can be separated in spatial averaged part $\overline{q}$ and
the rest quantity $Q(x,y,t)$:
\begin{equation}\label{eq7}
q(x,y,t) = \overline{q}(t) + Q(x,y,t)  .
\end{equation}
Let $\mid S \mid $ be large enough so that the plane average of any product of the rest
quantities vanish: $\overline{Q_i} = \overline{Q_i Q_j} = \overline{Q_i Q_j Q_k} =
\dots = 0 $ \footnote{The products $Q_i Q_j$; $Q_i Q_j Q_k$ have finite values and by definition of the averages very large
$\mid S \mid$ is presented in the denominator of the averaged quantity.}. In addition we shall assume that $\int \int _S dx \ dy \nabla^2 Q$ has
finite and small value such that $\overline{\nabla^2 Q} = \frac{1}{\mid S \mid} \int \int _S dx \ dy \nabla^2 Q
 \to 0$. The application of the averaging to Eq. (\ref{eq1}) in presence of
the assumptions given by Eqs. (\ref{eq2}), (\ref{eq3}) and (\ref{eq4}) (note
that in this case we have two spatial dimensions) leads to the system of
ODEs as follows ($i=1,2,\dots,N$):
\begin{equation}\label{eq8}
\frac{d \overline{\rho}_i}{d t} = \sum_{n_1=0}^\infty \sum_{n_2=0}^\infty \dots
\sum_{n_N=0}^\infty \alpha^{(i)}_{n_1,n_2,\dots,n_N}  
\overline{\rho}_1^{n_1} \overline{\rho}_{2}^{n_2} \dots \overline{\rho}_{N}^{n_N} .
\end{equation}
We note that Eq. (\ref{eq8}) follows directly from Eq. (\ref{eq5}) in the spatially homogeneous case.
The above discussion shows that Eq.(\ref{eq8}) can arise also in the spatially inhomogeneous case.
For the case of one population Eq.(\ref{eq8}) becomes
\begin{equation}\label{eq9}
\frac{d \overline{\rho}}{d t} = \sum_{n_1=0}^\infty  \alpha_{n_1} \overline{\rho}^{n_1} . 
\end{equation}
We note that the equations of the kind of Eqs.(\ref{eq8}) and (\ref{eq9}) are
often used as model equations in population dynamics not only for small values of
population densities but also for large values of these densities, i.e., for large
$\overline{\rho}_i$.
One example is connected to Holling functional response functions in predator-prey systems \cite{holling}.
For the case of one predator and one prey the functional response can be for an example type II
Holling functional response: $f(\overline{\rho})=\cfrac{a \overline{\rho}}{1+ a h \overline{\rho}}$ where 
$\overline{\rho}$ is the prey density
and $a$ and $h$ are parameters. The functional response can be also type III Holing functional response:
$f(\overline{\rho}) = \cfrac{\overline{\rho}^2}{h+ \overline{\rho}^2}$. The functional response function for the case of many prey species
can be more complicated. 
\par
For small values of population densities the models even with complicated nonlinear functional
responses can be reduced to models with polynomial nonlinearities. An example is
the one predator - two prey model \cite{leeuw}
\begin{eqnarray}\label{eq5d}
\frac{d \overline{\rho}_1}{dt} &=& \overline{\rho}_1 g_1(\overline{\rho}_1,\overline{\rho}_2) - 
\overline{\rho}_3 f_1(\overline{\rho}_1,\overline{\rho}_2)  \nonumber\\
\frac{d \overline{\rho}_2}{dt} &=& \overline{\rho}_2 g_2(\overline{\rho}_1,\overline{\rho}_2) - 
\overline{\rho}_3 f_2(\overline{\rho}_1,\overline{\rho}_2)  \nonumber\\
\frac{d \overline{\rho}_3}{dt} &=& \overline{\rho}_3 [c_1 f_1(\overline{\rho}_1,\overline{\rho}_2) + 
c_2 f_2(\overline{\rho}_1,\overline{\rho}_2)] - m \overline{\rho}_3  
\end{eqnarray}
where $f_i$ are the functional responses; $m$ is the constant mortality rate of the predator;
$c_i$ are the conversion factors of captured prey species into predators and $g_i$ are the growth
functions of the corresponding prey type. $\overline{\rho}_{1,2}$ are the spatial densities of the two types
of prey and $\overline{\rho}_3$ is the spatial density of the predator species. For small population
densities we can apply Taylor series expansion for the functions $f_{1,2}$ and $g_{1,2}$ and
thus obtain a system of equations from the studied in this paper class of equations
\begin{eqnarray}\label{eq5e}
\frac{d \overline{\rho}_1}{dt}&=& \sum_{n_1=0}^{\infty} \sum_{n_2=0}^{\infty} \frac{\overline{\rho}_1^{n_1} \overline{\rho}_2^{n_2}}{n_1! n_2!}
(\overline{\rho}_1 \alpha_{3,n_1,n_2} - \overline{\rho}_3 \alpha_{1,n_1,n_2}) \nonumber \\
\frac{d \overline{\rho}_2}{dt}&=& \sum_{n_1=0}^{\infty} \sum_{n_2=0}^{\infty} \frac{\overline{\rho}_1^{n_1} \overline{\rho}_2^{n_2}}{n_1! n_2!}
(\overline{\rho}_2 \alpha_{4,n_1,n_2} - \overline{\rho}_3 \alpha_{2,n_1,n_2}) \nonumber \\
\frac{d \overline{\rho}_3}{dt}&=& \overline{\rho}_3 \bigg[ - m + 
\sum_{n_1=0}^{\infty} \sum_{n_2=0}^{\infty} \frac{\overline{\rho}_1^{n_1} \overline{\rho}_2^{n_2}}{n_1! n_2!}
(c_1 \alpha_{1,n_1,n_2} + c_2 \alpha_{2,n_1,n_2}) \bigg] \nonumber \\
\end{eqnarray}
where
\begin{eqnarray*}
\alpha_{1,n_1,n_2} = \left( \frac{\partial^{n_1+n_2} f_1}{\partial \rho_1^{n_1} \partial \rho_2^{n_2}}\right) \Bigg 
\vert_{\overline{\rho}_1=\overline{\rho}_2=0}; \
\alpha_{2,n_1,n_2} = \left( \frac{\partial^{n_1+n_2} f_2}{\partial \rho_1^{n_1} \partial \rho_2^{n_2}}\right) \Bigg 
 \vert_{\overline{\rho}_1=\overline{\rho}_2=0};\ \\
\alpha_{3,n_1,n_2} = \left( \frac{\partial^{n_1+n_2} g_1}{\partial \rho_1^{n_1} \partial \rho_2^{n_2}}\right) 
\vert_{\overline{\rho}_1=\overline{\rho}_2=0}; \
\alpha_{4,n_1,n_2} = \left( \frac{\partial^{n_1+n_2} g_2}{\partial \rho_1^{n_1} \partial \rho_2^{n_2}}\right) \Bigg 
\vert_{\overline{\rho}_1=\overline{\rho}_2=0}; \
\end{eqnarray*} 
\section{Traveling waves}
\subsection{Case of one population}
Let us discuss the simplest case of one population  described by Eq. (\ref{eq5a}). First we
introduce the traveling-wave coordinate $\xi = x-vt$ where $v$ is the velocity
of the wave. In addition we shall assume that the polynomial nonlinearity in 
Eq.(\ref{eq5a}) is up to order $L$. We rescale the coefficients in Eq.(\ref{eq5a})
as follows: 
\begin{equation}\label{et0}
D^\dag = -D/v; \ \alpha_{n_1}^\dag = \alpha_{n_1}/v. 
\end{equation}
Then Eq.(\ref{eq5a}) becomes:
\begin{equation}\label{et1}
\frac{d \overline{\rho}}{d \xi} + D^\dag \frac{d^2 \overline{\rho}}{d \xi^2} + \sum_{n_1=0}^{L} \alpha_{n_1}^\dag
\overline{\rho}^{n_1} =0 .
\end{equation}
Below we shall obtain exact solution  of Eq. (\ref{et1}) by application of the 
modified method of simplest equation for obtaining exact solutions of nonlinear PDEs.
For more details on the modified method of simplest equation see Appendix A.
\newtheorem*{prop1}{Proposition 1}
\begin{prop1}
The balance equation for Eq. (\ref{et1}) for the case when Riccati equation is
used as simplest equation is $P(L-1) = 2$ where $P$ is the largest power in the
polynomial for $\overline{\rho}(\xi)$ constructed on the basis of the solutions $\Phi(\xi)$
of the Riccati equation.
\end{prop1}
\begin{proof}
\par
We apply the methodology from Appendix A to Eq.(\ref{et1}).  We constrict  a solution as
finite series
\begin{equation}\label{et6}
\rho(\xi) = \sum_{i=0}^P a_i [\Phi(\xi)]^i,
\end{equation}
where $\Phi(\xi)$ is a solution of the Riccati equation
\begin{equation}\label{et7}
\frac{d \Phi}{d \xi} = a \phi^2 + b \phi + c,
\end{equation}
i.e.,
\begin{equation}\label{et8}
\Phi(\xi) = - \frac{b}{2a} - \frac{\theta}{2a} \tanh \left[\frac{\theta (\xi + \xi_0)}{2} \right] ,\  \theta^2 = b^2 - 4 ac
\end{equation}
The substitution of Eq.(\ref{et6}) in Eq.(\ref{et1}) and the balance of the largest powers
of $\Phi$ that arise from the different terms of Eq.(\ref{et1}) (these powers are $P+2$ from the term 
$\cfrac{d^2 \overline{\rho}}{d \xi^2}$ and $PL$ from the term $\alpha_{L}^\dag
\overline{\rho}^{L}$) lead to the balance equation
\begin{equation}\label{et9}
P(L-1)=2 .
\end{equation}
\end{proof}
Thus we have the possibilities: $P=L=2$ or $P=1;L=3$. Below we discuss these possibilities.
\subsection{Case $P=L=2$}
 In this case  we shall obtain exact traveling-wave solution of the equation
\begin{equation}\label{et10}
\frac{\partial \rho}{\partial t} - D \frac{\partial^2 \rho}{\partial x^2} =
\alpha_0 + \alpha_1 \rho + \alpha_2 \rho^2 ,
\end{equation}
\newtheorem*{prop2}{Proposition 2}
\begin{prop2}
The traveling wave solution of the kind Eq.(\ref{et6}) of Eq. (\ref{et10}) obtained on the basis 
of the modified method of simplest equation when the equation of Riccati is used as simplest equation is
\begin{eqnarray}\label{ss1}
\overline{\rho}(\xi) = \frac{75 {D^\dag}^2 b^2 + 30 {D^\dag} b - 3 + 25 \alpha_1^\dag D^\dag}{50 
\alpha_2^\dag D^\dag} +
\frac{3[(25 {D^\dag}^2 b^2 - 1)(5 {D^\dag}  b + 1)]}{250 \alpha_2^\dag c {D^\dag}^2} \times \nonumber \\ 
\left \{  \frac{b}{2\frac{25 {D^\dag}^2 b^2-1}{100 c {D^\dag}^2}} + \frac{\theta}{2\frac{25 {D^\dag}^2 b^2-1}{100 c {D^\dag}^2}} \tanh \left[\frac{\theta (\xi + 
\xi_0)}{2} \right]\right \} 
 - \frac{3 (25 {D^\dag}^2 b^2-1)}{5000 \alpha_2^\dag c^2 {D^\dag}^3} \times \nonumber \\
\left \{  \frac{b}{2\frac{25 {D^\dag}^2 b^2-1}{100 c {D^\dag}^2}}  +
\frac{\theta}{2\frac{25 {D^\dag}^2 b^2-1}{100 c {D^\dag}^2}} \tanh \left[\frac{\theta (\xi + \xi_0)}{2} \right]\right \}^2,  \nonumber \\
\theta^2 = b^2 - \frac{25 {D^\dag}^2 b^2-1}{25 {D^\dag}^2},
\end{eqnarray}
for the case $\alpha_0^\dag \ne 0$ and
\begin{eqnarray}\label{ss2}
\overline{\rho}(\xi) = - \frac{36 b^2 + 60 b \alpha_1^\dag + 25 {\alpha_1^\dag}^2}{100 \alpha_1^\dag \alpha_2^\dag} +
\frac{(36 b^2 - 25 {\alpha_1^\dag}^2)(6 b + 5 \alpha_1^\dag)}{600 c \alpha_1^\dag \alpha_2^\dag} \Bigg \{  \frac{b}{2a} + 
\nonumber \\ 
 \frac{\theta}{2a} \tanh \left[\frac{\theta (\xi + 
\xi_0)}{2} \right]\Bigg \} - 
 \frac{(36 b^2 - 25 {\alpha_1^\dag}^2)^2}{14400 c \alpha_1^\dag \alpha_2^\dag} \left \{  \frac{b}{2a}  +
\frac{\theta}{2a} \tanh \left[\frac{\theta (\xi + \xi_0)}{2} \right]\right \}^2 , \nonumber \\
\theta^2 = b^2 - \frac{36 b^2 - 25 {\alpha_1^\dag}^2}{36} 
\end{eqnarray}
for the case $\alpha_0^\dag = 0$. $\xi_0$ is a constant of integration.
\end{prop2}
\begin{proof}
As $P=L=2$ then from Eq.(\ref{et6}) the solution will be of the kind
\begin{eqnarray}\label{et11}
\overline{\rho}(\xi) = a_0 - a_1 \left \{  \frac{b}{2a} + \frac{\theta}{2a} \tanh \left[\frac{\theta (\xi + 
\xi_0)}{2} \right]\right \} + 
a_2 \left \{  \frac{b}{2a}  +
\frac{\theta}{2a} \tanh \left[\frac{\theta (\xi + \xi_0)}{2} \right]\right \}^2 .
\end{eqnarray}
The substitution of Eq.(\ref{et11}) in Eq.(\ref{et10}) leads to a system of relationships for 
the parameters of the solution (the  system is of kind (\ref{et5})).
\begin{eqnarray}\label{et12}
6 D^\dag a_2 a^2 + \alpha_2^\dag a_2^2 &=&0 , \nonumber \\
a D^\dag (a_1 a + 5 a_2  b) + a_2 a +  \alpha_2^\dag a_1 a_2 &=&0 , \nonumber\\
D^\dag [3 a_1 a b + 4 a_2 (2 a c + b^2)] + a_1 a + 
2 a_2 b + \alpha_1^\dag a_2 + \alpha_2^\dag (2 a_0 a_2 + a_1^2) &=&0 , \nonumber\\
D^\dag [a_1 (2 a c + b^2) + 6 a_2 b c] + \alpha_1^\dag a_1 +
2 \alpha_2^\dag a_0 a_1 + a_1 b + 2 a_2 c &=& 0 , \nonumber \\
\alpha_0^\dag + \alpha_1^\dag a_0 + \alpha_2^\dag a_0^2 + a_1 c + D^\dag  (a_1 b c + 2 a_2 c^2) &=& 0 . 
\nonumber \\
\end{eqnarray}
Now we have 2 possibilities: $\alpha_0^\dag \ne 0$ and $\alpha_0^\dag =0$ (which is more close to the
classical population dynamics models that usually do not possess terms independent on the
population density).
\begin{flushleft}
\emph{Case $\alpha_0^\dag \ne 0$}
\end{flushleft}
For this case the solution of the system (\ref{et12}) is as follows:
\begin{eqnarray}\label{et13}
\alpha_0^\dag &=& \frac{625 {\alpha_1^\dag}^2 {D^\dag}^2 - 36}{2500 \alpha_2^\dag {D^\dag}^2};\ 
a_0 = - \frac{75 {D^\dag}^2 b^2 + 30 {D^\dag} b - 3 + 25 \alpha_1^\dag D^\dag}{50 \alpha_2^\dag D^\dag},
\nonumber \\
a_1 &=& -\frac{3[(25 {D^\dag}^2 b^2 - 1)(5 {D^\dag}  b + 1)]}{250 \alpha_2^\dag c D^\dag};\ 
a_2 = - \frac{3 (25 {D^\dag}^2 b^2-1)}{5000 \alpha_2^\dag c^2 {D^\dag}^3}, \nonumber \\
a &=& \frac{25 {D^\dag}^2 b^2-1}{100 c {D^\dag}^2} .
\end{eqnarray}
and then we obtain the solution (\ref{ss1}) of Eq.(\ref{et10})
\begin{flushleft}
\emph{Case $\alpha_0^\dag = 0$}
\end{flushleft}
For this case the solution of the system (\ref{et12}) is as follows:
\begin{eqnarray}\label{et14}
D^\dag &=& \frac{6}{25 \alpha_1^\dag} ; \
a_0 = - \frac{36 b^2 + 60 b \alpha_1^\dag + 25 {\alpha_1^\dag}^2}{100 \alpha_1^\dag \alpha_2^\dag} ,
\nonumber \\
a_1 &=& -\frac{(36 b^2 - 25 {\alpha_1^\dag}^2)(6 b + 5 \alpha_1^\dag)}{600 c \alpha_1^\dag
\alpha_2^\dag} ; 
a_2 = - \frac{(36 b^2 - 25 {\alpha_1^\dag}^2)^2}{14400 c \alpha_1^\dag \alpha_2^\dag} , \nonumber \\
a &=& \frac{36 b^2 - 25 {\alpha_1^\dag}^2}{144 c} .
\end{eqnarray}
and then we obtain the solution (\ref{ss2}) of Eq.(\ref{et10}). 
\end{proof}
The obtained solutions describe kink waves that can be considered as traveling waves of
change of the value of the population density of the studied population. Such waves
can describe the front of migration of a population. The appropriate values of the
boundary conditions ensure that $\rho(\xi)$ is non-negative elsewhere. We note that the
parameters of the solved Eq.(\ref{et10}) are $D^\dag$ and $\alpha_0^\dag, \alpha_1^\dag,
\alpha_2^\dag$. The first relationship from Eq. (\ref{et13}) connects these 4 parameters.
Then the solution (\ref{ss1}) does not hold for any values of parameters of Eq.(\ref{et10})
but only for these combinations that satisfy the above mentioned relationship.
\par
Let us discuss in several words the problem with the boundary conditions of the
obtained solutions. theoretically for the general case of solution (\ref{ss1}) there are 10 
parameters and 5 relationships (\ref{et12}) among them. Thus there are 5 free parameters
and several possibilities for boundary conditions are
\begin{eqnarray}\label{et15}
\overline{\rho}(+\infty)&=&A_1; \ \overline{\rho}(-\infty)=A_2; \nonumber \\
\frac{d \overline{\rho}}{d \xi} \bigg \vert_{B_1} &=& A_3, \ (B_1>0); \
\frac{d \overline{\rho}}{d \xi} \bigg \vert_{-B_2} = A_4, \ (B_2 >0); \nonumber \\
\frac{d^2 \overline{\rho}}{d \xi^2} \bigg \vert_{+B_3} &=& A_5, \ (B_3 >0) \
\frac{d^2 \overline{\rho}}{d \xi^2} \bigg \vert_{-B_4} = A_6, \ (B_4 >0); \nonumber \\
\dots && 
\end{eqnarray}
Let us impose the boundary conditions $\overline{\rho}(+\infty)= A_1; \ 
\overline{\rho}(-\infty)=A_2$ on the solution given by Eq.(\ref{ss1}). The
result is that there are two additional relationships that must be
satisfied by the  parameters of the solution. The relationships are as follows
\begin{eqnarray}\label{et16}
D^\dag = \frac{6}{25}\frac{1}{ \alpha_1^\dag + 2 A_1 \alpha_2^\dag};\
\alpha_1^\dag = - \alpha_2^\dag (A_1 + A_2)
\end{eqnarray}
The second relationship can  be written as follows: $A_2 = - (A_1 + 
\alpha_1^\dag/\alpha_2^\dag)$ which means that if the parameters $\alpha_{1,2}^\dag$
of the solved PDE are fixed then when we choose the boundary condition $A_1$
then the boundary condition $A_2$ can not be arbitrary. For an example if $A_1=0$
then $A_2 = - \alpha_1^\dag/\alpha_2^\dag$.
\begin{figure}[t]
\begin{center}
\includegraphics[scale=0.8]{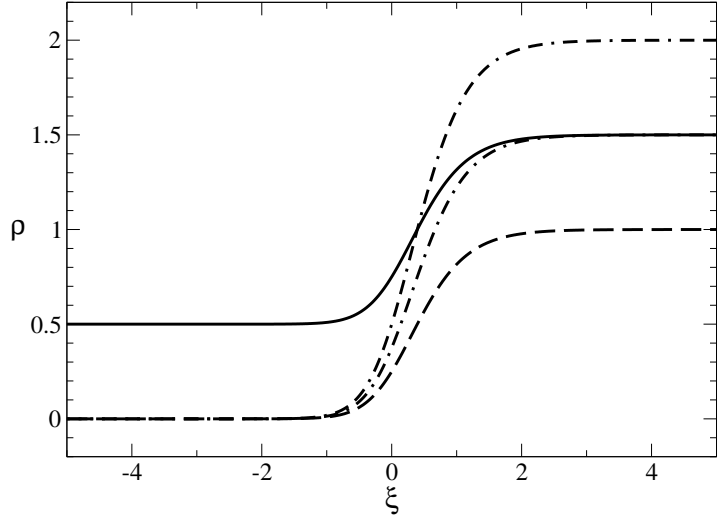}
\end{center}
\caption{Several solutions that satisfy boundary conditions $\overline{\rho}(+\infty)= A_1; 
\ \overline{\rho}(-\infty)=A_2$. Solid line: $A_1=1.5$; $A_2=0.5$. Dashed line:
$A_1=1$; $A_2=0$. Dot-dashed line: $A_1=1.5$; $A_2=0$. Dot-double-dashed line:
$A_1=0$; $A_2=2$.}
\end{figure}
Several examples of the obtained nonlinear waves satisfying two boundary conditions
are shown in Fig.1.
\subsubsection{Case $P=1$; $L=3$}
In this case we shall obtain exact traveling-wave solutions  of the equation
\begin{equation}\label{s1}
\frac{\partial \rho}{\partial t} - D \frac{\partial^2 \rho}{\partial x^2} =
\alpha_0 + \alpha_1 \rho + \alpha_2 \rho^2 + \alpha_3 \rho^3,
\end{equation}
\newtheorem*{prop2a}{Proposition 3}
\begin{prop2a}
The traveling wave solution of the kind Eq.(\ref{et6}) of Eq. (\ref{s1}) obtained on the basis 
of the modified method of simplest equation when the equation of Riccati is used as simplest equation is
\begin{eqnarray}\label{s4}
\overline{\rho}(\xi) = a_0 - a_1 \left \{  \frac{b}{2a} + \frac{\theta}{2a} \tanh \left[\frac{\theta (\xi + 
\xi_0)}{2} \right]\right \}
\end{eqnarray}
for the equation
\begin{eqnarray}\label{s4a}
\frac{\partial \rho}{\partial t} - D \frac{\partial^2 \rho}{\partial x^2} = \nonumber \\
- v\frac{-3 a a_0^2 D^\dag a_1 b + 2 D^\dag a^2 a_0^3 - a a_0^2 a_1 - a_1^3 c + a_0 a_1^2 b + 
2 a_0 D^\dag a_1^2 a c + a_0 D^\dag a_1^2 b^2 - D^\dag a_1^3 b c}{a_1^2} - \nonumber \\  
 v \frac{a_1^2 b + 2 D^\dag a_1^2 a c + D^\dag a_1^2 b^2 - 6 a a_0 D^\dag a_1 b + 
6 D^\dag a^2 a_0^2 - 2 a a_0 a_1}{a_1^2} \rho + \nonumber \\
v \frac{a(-3 D^\dag a_1 b + 6 D^\dag a a_0 - a_1)}{a_1^2} \rho^2 
 -2 v \frac{D^\dag a^2}{a_1^2} \rho^3 \nonumber \\
\end{eqnarray}
and \\
\begin{equation}\label{s4b}
\rho(\xi) = \frac{6 D^\dag (-2 \alpha_3^\dag D^\dag)^{1/2} b - 2 \alpha_2^\dag D^\dag + (-2 \alpha_3^\dag D^\dag)^{1/2}+
3 (-2 \alpha_3^\dag D^\dag)^{1/2} D^\dag \theta \tanh \left[ \cfrac{\theta (\xi+\xi_0)}{2} \right]}{6 \alpha_3^\dag D^\dag} 
\end{equation}
for the equation 
\begin{eqnarray}\label{s4c}
&&\frac{\partial \rho}{\partial t} - D \frac{\partial^2 \rho}{\partial x^2} =
-\frac{v}{216 {\alpha_3^\dag}^2 {D^\dag}^3} 
\bigg[ 9 b D^\dag (-2 \alpha_3^\dag D^\dag)^{3/2} +
27 {D^\dag}^2 (-2 \alpha_3^\dag D^\dag)^{3/2} b^2 + \nonumber \\
&& 27 {D^\dag}^3 (-2 \alpha_3^\dag D^\dag)^{3/2} b^3 +
18 (-2 \alpha_3^\dag D^\dag)^{1/2} D^\dag \alpha_3 + 24 \alpha_2^2 {D^\dag}^2 (-2 \alpha_3 D^\dag)^{1/2}+ \nonumber \\
&& 54 (-2 \alpha_3 D^\dag)^{1/2} {D^\dag}^3 \alpha_3 b^2 - 72 (-2 \alpha_3 D^\dag)^{1/2} {D^dag}^2 \alpha_1^\dag \alpha_3^\dag +
16 {\alpha_2^\dag}^3 {D^\dag}^3 + \nonumber \\
&& 18 (-2 \alpha_3^\dag D^\dag)^{1/2} b {D^\dag}^2 \alpha_3^\dag+ 
54 (-2 \alpha_3^\dag D^\dag)^{1/2} b^3 {D^\dag}^4 \alpha_3^\dag+
(-2 \alpha_3^\dag D^\dag)^{3/2} - \nonumber \\
&& 72 \alpha_1^\dag \alpha_3^\dag {D^\dag}^3 \alpha_2) \bigg]
+ \alpha_1 \rho + \alpha_2 \rho^2 + \alpha_3 \rho^3,
\end{eqnarray}
\end{prop2a}
\begin{proof}
As $P=1$;$L=3$ then from Eq.(\ref{et6}) the solution will be of the kind Eq.(\ref{s4})
The substitution of Eq.(\ref{s4}) in Eq.(\ref{s1}) leads to the following
system of relationships for the parameters of the solution
\begin{eqnarray}\label{s5}
2 D^\dag a_1 a^2 + \alpha_3^\dag a_1^3 &=& 0 \nonumber \\
a_1(3 D^\dag  a b + \alpha_2^\dag a_1 + 3 \alpha_3^\dag a_0 a_1 +  a) &=& 0 \nonumber \\
a_1 b + D^\dag a_1 (2 a c + b^2) + 2 \alpha_2^\dag a_0 a_1 + \alpha_1^\dag a_1 + 3 \alpha_3^\dag a_0^2 a_1 &=& 0 \nonumber \\
\alpha_2^\dag a_0^2 + a_1 c + \alpha_1^\dag a_0 + \alpha_3^\dag a_0^3 + D^\dag a_1 b c + \alpha_0^\dag &=& 0
\end{eqnarray}
We shall consider two possibilities. First we shall solve the system (\ref{s5}) with respect to 
$\alpha_{0,1,2,3}^\dag$. This means that the coefficients $a_{0,1}$ as well as the coefficients
$a,b,c$ of the Riccati equation will remain free and we can impose many boundary conditions
on the solution. Thus we shall investigate a small subclass of equation of kind Eq.(\ref{s1})
where we have solution with many free parameters. Then we shall consider the case in
which we shall keep as many as possible of the parameters of Eq.(\ref{s1}) free. The price
for this will be that we shall have to impose several relationships on the coefficients of
the solutions\\
\emph{Case 1: Solution of (\ref{s5}) with respect to $\alpha_{0,1,2,3}^\dag$}\\
This solution is as follows
\begin{eqnarray}\label{s6}
\alpha_0^\dag &=& - \frac{-3 a a_0^2 D^\dag a_1 b + 2 D^\dag a^2 a_0^3 - a a_0^2 a_1 - a_1^3 c + a_0 a_1^2 b + 
2 a_0 D^\dag a_1^2 a c + a_0 D^\dag a_1^2 b^2 - D^\dag a_1^3 b c}{a_1^2} \nonumber \\
\alpha_1^\dag &=& - \frac{a_1^2 b + 2 D^\dag a_1^2 a c + D^\dag a_1^2 b^2 - 6 a a_0 D^\dag a_1 b + 
6 D^\dag a^2 a_0^2 - 2 a a_0 a_1}{a_1^2} \nonumber \\
\alpha_2^\dag &=& \frac{a(-3 D^\dag a_1 b + 6 D^\dag a a_0 - a_1)}{a_1^2} \nonumber \\
\alpha_3^\dag &=& -2 \frac{D^\dag a^2}{a_1^2} \nonumber\\
\end{eqnarray}
Thus we arrive at solution (\ref{s4}) of Eq. (\ref{s4a}).\\
\emph{Case 2: Solution of (\ref{s5}) with respect to $a,a_0,a_1$ and $\alpha_0^\dag$} \\
The solution is as follows
\begin{eqnarray}\label{s7}
a&=& \frac{3 {D^\dag}^2 \alpha_3 b^2 + 2 {\alpha_2^\dag}^2 D^\dag - 6 \alpha_1^\dag \alpha_3^\dag D^\dag + \alpha_3^\dag
}{12 c {D^\dag}^2 \alpha_3^\dag} \nonumber \\
a_0 &=& \frac{(3 D^\dag b + 1) \sqrt{-2 \alpha_3^\dag D^\dag} - 2 \alpha_2 D^\dag
}{6 D^\dag \alpha_3^\dag} \nonumber \\
a_1 &=& \frac{\sqrt{-2 \alpha_3^\dag D^\dag}(3 {D^\dag}^2 \alpha_3^\dag b^2 + 2 \alpha_2^2 D^\dag - 
6 \alpha_1^\dag \alpha_3 D^\dag + \alpha_3^\dag)}{12 {\alpha_3^\dag}^2 c {D^\dag}^2}\nonumber \\
\alpha_0^\dag &=& - \frac{1}{216 {\alpha_3^\dag}^2 {D^\dag}^3} 
\bigg[ 9 b D^\dag (-2 \alpha_3^\dag D^\dag)^{3/2} +
27 {D^\dag}^2 (-2 \alpha_3^\dag D^\dag)^{3/2} b^2 + \nonumber \\
&& 27 {D^\dag}^3 (-2 \alpha_3^\dag D^\dag)^{3/2} b^3 +
18 (-2 \alpha_3^\dag D^\dag)^{1/2} D^\dag \alpha_3 + 24 \alpha_2^2 {D^\dag}^2 (-2 \alpha_3 D^\dag)^{1/2}+ \nonumber \\
&& 54 (-2 \alpha_3 D^\dag)^{1/2} {D^\dag}^3 \alpha_3 b^2 - 72 (-2 \alpha_3 D^\dag)^{1/2} {D^dag}^2 \alpha_1^\dag \alpha_3^\dag +
16 {\alpha_2^\dag}^3 {D^\dag}^3 + \nonumber \\
&& 18 (-2 \alpha_3^\dag D^\dag)^{1/2} b {D^\dag}^2 \alpha_3^\dag+ 
54 (-2 \alpha_3^\dag D^\dag)^{1/2} b^3 {D^\dag}^4 \alpha_3^\dag+
(-2 \alpha_3^\dag D^\dag)^{3/2} - \nonumber \\
&& 72 \alpha_1^\dag \alpha_3^\dag {D^\dag}^3 \alpha_2) \bigg]
\nonumber \\
\end{eqnarray}
Thus we arrive at solution (\ref{s4b}) of Eq. (\ref{s4c}).
\end{proof}
\par
We can now impose boundary conditions of the solution (\ref{s4}) of Eq. (\ref{s4a}). For an
example the boundary conditions can be 
\begin{eqnarray*}\label{s*}
\overline{\rho}(+\infty)&=&A_1; \ \overline{\rho}(-\infty)=A_2; \nonumber \\
\frac{d \overline{\rho}}{d \xi} \bigg \vert_{B_1} &=& A_3, \ (B_1>0); \
\frac{d \overline{\rho}}{d \xi} \bigg \vert_{-B_2} = A_4, \ (B_2 >0) \nonumber \\
\end{eqnarray*}
The boundary conditions will fix additional parameters of the solution.
We let the corresponding algebraic manipulations to the interested reader.
\subsection{Coupled waves in a system of 3 populations}
Let us discuss a system of 3 competing populations modeled by the system of Lotka-Volterra kind
(\ref{eq5b}) for the case of constant coefficients of change of population members and constant
interaction coefficients, i.e., for the case $r_{ik}=0$; $\alpha_{ijk}=0$. The system of equations
becomes
\begin{eqnarray}\label{d1}
\frac{\partial \overline{\rho}_1}{\partial t} -D_{11}\frac{\partial \overline{\rho}_1}{\partial
x}-D_{12}\frac{\partial \overline{\rho}_2}{\partial x} -
D_{13}\frac{\partial \overline{\rho}_3}{\partial x}= r_1^0 \overline{\rho}_1 - r_1^0
\alpha_{11}^0 \overline{\rho}_1^2 - r_1^0 \alpha_{12}^0 \overline{\rho}_1 \overline{\rho}_2 -
r_1^0 \alpha_{13}^0 \overline{\rho}_1 \overline{\rho}_3
\nonumber \\
\frac{\partial \overline{\rho}_2}{\partial t} -D_{21}\frac{\partial \overline{\rho}_1}{\partial
x}-D_{22}\frac{\partial \overline{\rho}_2}{\partial x} -
D_{23}\frac{\partial \overline{\rho}_3}{\partial x}= r_2^0 \overline{\rho}_2 - r_2^0
\alpha_{21}^0 \overline{\rho}_1 \overline{\rho}_2 - r_2^0 \alpha_{22}^0  \overline{\rho}_2^2 -
r_2^0 \alpha_{23}^0 \overline{\rho}_2 \overline{\rho}_3
\nonumber \\
\frac{\partial \overline{\rho}_3}{\partial t} -D_{31}\frac{\partial \overline{\rho}_1}{\partial
x}-D_{32}\frac{\partial \overline{\rho}_2}{\partial x} -
D_{33}\frac{\partial \overline{\rho}_3}{\partial x}= r_3^0 \overline{\rho}_3 - r_3^0
\alpha_{31}^0 \overline{\rho}_1 \overline{\rho}_3 - r_3^0 \alpha_{32}^0  \overline{\rho}_2
\overline{\rho}_3 - r_3^0 \alpha_{33}^0  \overline{\rho}_3^2 \nonumber \\
\end{eqnarray}
For this system we shall demonstrate existence of a simple coupled kink wave solution (more complicated
solutions are possible too).
\newtheorem*{prop4}{Proposition 4}
\begin{prop4}
The system (\ref{d1}) possesses coupled kind wave solution of the kind
\begin{eqnarray}\label{d1s}
\overline{\rho}_1(\xi) &=& -\frac{a_1 c_0}{a_1 + b_1} + a_1 
\left \{ \frac{a_1 + b_1 + 4 a_1 c_0}{4 a_1  (a_1 + b_1)} + 
\frac{\theta (a_1 + b_1 + 4 a_1 c_0)}{4 a_1 b (a_1 + b_1)} \tanh \left[\frac{\theta (\xi + \xi_0)}{2} \right] \right \}, \nonumber\\
\overline{\rho}_2(\xi) &=& - \frac{b_1 c_0}{a_1 + b_1} + b_1 \left \{ 
\frac{a_1 + b_1 + 4 a_1 c_0}{4 a_1  (a_1 + b_1)} + 
\frac{\theta (a_1 + b_1 + 4 a_1 c_0)}{4 a_1 b (a_1 + b_1)} \tanh \left[\frac{\theta (\xi + \xi_0)}{2} \right]
\right \}, \nonumber \\
\overline{\rho}_3(\xi) &=& -(a_1+b_1) + c_1 \left \{ 
\frac{a_1 + b_1 + 4 a_1 c_0}{4 a_1  (a_1 + b_1)} + 
\frac{\theta (a_1 + b_1 + 4 a_1 c_0)}{4 a_1 b (a_1 + b_1)} \tanh \left[\frac{\theta (\xi + \xi_0)}{2} \right]
 \right \}, \nonumber\\
\end{eqnarray}
where $\xi - x + \cfrac{4 a_1 c_0 + (a_1+b_1)(1-b+ b D_{11})}{b(a_1+b_1)} t$. $a_1$, $b$, $b_1$, $c_0$, $D_{11}$
are free parameters. 
\end{prop4}
We remember that above $\theta^2 = b^2 - 4 ac$ where $a, b, c$ are the parameters of the Riccati equation.
\begin{proof}
We apply the modified method of simplest equation to the system (\ref{d1}).
First of all we introduce the traveling wave coordinate $\xi = x-vt$ where $v$ is the wave velocity.
Then we search for the solution in the form
\begin{eqnarray}\label{d2}
\overline{\rho}_1(\xi) = \sum_{i=0}^{P} a_i \Phi(\xi)^i; \ \overline{\rho}_2(\xi) = \sum_{j=0}^{Q} b_j \Phi(\xi)^j;
\ \overline{\rho}_3(\xi) = \sum_{k=0}^{R} c_k \Phi(\xi)^k
\end{eqnarray}
where $\cfrac{d \Phi}{d \xi} = a \Phi^2 + b \Phi + c$. The simplest possible balance equation is $P=Q=R=1$. The substitution of all
above in the system (\ref{d1}) leads to the following  system of 9 nonlinear algebraic equations
\begin{eqnarray}\label{d3}
1.) && - D_{13}c_1a -(v+D_{11})a_1a -r_{10} \alpha_{120}a_1b_1 + r_{10} \alpha_{110} a_1^2 - D_{12} b_1 a - r_{10} \alpha_{130} a_1 c_1 = 0 
\nonumber \\
2.) &&- r_{10} \alpha_{120} a_0 b_1 - r_{10} \alpha_{120} a_1 b_0 + 2 r_{10} \alpha_{110} a_0 a_1 - r_{10} \alpha_{130} a_0 c_1 - 
 r_{10} \alpha_{130} a_1 c_0 - r_{10} a_1 - \nonumber \\
&& D_{13} c_1 b - D_{12} b_1 b -(v + D_{11}) a_1 b =0 
\nonumber \\
3.) && -(v + D_{11})a_1 c - r_{10} \alpha_{120} a_0 b_0 - D_{13} c_1 c - D_{12} b_1 c - r_{10} \alpha_{130} a_0 c_0 - r_{10} a_0 +\nonumber \\
&& r_{10} \alpha_{110} a_0^2 =0
\nonumber \\
4.) && -D_{23} c_1 a - D_{21} a_1 a - r_{20} \alpha_{220} b_1^2 + r_{20} \alpha_{210} a_1 b_1 - (v + D_{22}) b_1 a - r_{20} \alpha_{230} b_1 c_1 = 0 
\nonumber \\
5.) && -2 r_{20} \alpha_{220} b_0 b_1 + r_{20} \alpha_{210} a_0 b_1 + r_{20} \alpha_{210} a_1 b_0 - r_{20} \alpha_{230} b_0 c_1 - 
  r_{20} \alpha_{230} b_1 c_0 - r_{20} b_1 - \nonumber \\
&& D_{23} c_1 b - (v+D_{22}) b_1 b - D_{21} a_1b = 0 \nonumber \\
6.) && -D_{21} a_1 c - r_{20} \alpha_{220} b_0^2 - D_{23} c_1 c -(v+D_{22}) b_1 c - r_{20} \alpha_{230} b_0 c_0 - r_{20} b_0 + r_{20} \alpha_{210} a_0 b_0 =0
\nonumber\\
7.) && -(v + D_{33}) c_1 a - D_{31} a_1 a - r_{30} \alpha_{320} b_1 c_1 + r_{30} \alpha_{310} a_1 c_1 - D_{32} b_1 a - r_{30} \alpha_{330} c_1^2 =0
\nonumber\\
8.) && -r_{30} \alpha_{320} b_0 c_1 - r_{30} \alpha_{320} b_1 c_0 + r_{30} \alpha_{310} a_0 c_1 + r_{30} \alpha_{310} a_1 c_0 - 2 r_{30} \alpha_{330} c_0 c_1 - r_{30} c_1 -\nonumber \\
&&  (v+D_{33}) c_1 b - D_{32} b_1 b - D_{31} a_1 b=0
\nonumber \\
9.) && -D_{31} a_1 c - r_{30} \alpha_{320} b_0 c_0 - (v+D_{33}) c_1 c - D_{32} b_1 c - r_{30} \alpha_{330} c_0^2 - r_{30} c_0 + r_{30} \alpha_{310} a_0 c_0=0
\nonumber \\
\end{eqnarray}
The general solution of this system is very long. In order to obtain  the solution from the text of Proposition 4
we fix some of the parameters in the above system as follows
\begin{eqnarray}\label{d3a}
r_{10}&=&1;r_{20}=1;r_{30}=1;\alpha_{110}=1;\alpha_{220}=1;\alpha_{330}=1;\alpha_{120}=1;\alpha_{130}=1;\alpha_{210}=1;\nonumber \\ 
\alpha_{230}&=&1; \alpha_{310}=1;\alpha_{320}=1;D_{21}=D_{12};D_{31}=D_{13};D_{32}=D_{23};D_{22}=D_{11}; D_{33}=D_{11}; \nonumber \\
D_{12} &=& 1;D_{13}=1;D_{23}=1. \nonumber \\
\end{eqnarray} 
In this simple case the solution of the system (\ref{d3}) is (for details of solution of the system of
equations by means of a Maple program see Appendix B)
\begin{eqnarray}\label{d4}
c_1 &=& -(a_1+b_1) \nonumber \\
v &=& - \frac{4 a_1 c_0 + b D_{11} a_1 - a_1 b + a_1-b_1 b + b D_{11} b_1 + b_1}{b(a_1+b_1)} \nonumber \\
c &=& - \frac{c_0 (2 a_1 c_0 + a_1 + b_1) b}{(a_1+b_1)(a_1+b_1+4 a_1 c_0)} \nonumber\\
a &=& -2\frac{a_1 b (a_1 + b_1)}{(a_1 + b_1 + 4 a_1 c_0)} \nonumber\\
a_0 &=& -\frac{a_1 c_0}{a_1 + b_1} \nonumber \\
b_0 &=& - \frac{b_1 c_0}{a_1 + b_1}
\end{eqnarray}
Substituting the coefficients in the functions $\overline{\rho_1}$, $\overline{\rho_2}$, $\overline{\rho_3}$
we obtain the coupled kink wave solution from the formulation of the Proposition 4.
\end{proof}
\section{Statistical distributions and exit time}
Eqs.(\ref{eq8}) and (\ref{eq9}) are typical equations for
description of the evolution of dynamical systems. The general case
of such equations is
\begin{equation}\label{ed1}
\frac{d x_i}{d t} = X_i(x_1,x_2,\dots,x_N); \ i=1,2,\dots,N.
\end{equation} 
where $X_i(x_1,x_2,\dots,x_N)$ is some (in the general case nonlinear) function.
For such kind of systems there exists a theory that allows us to
characterize some system properties in the case when the system
is under the action of random perturbations. Pontryagin, Andronov and Vitt \cite{pav}
developed such theory for random impulses that occur after every
interval of time $\tau$ and each impulse causes the phase point of
the dynamical system described by Eqs. (\ref{ed1}) to jump
through a distance $a$ along a random direction. Let us first
consider the case of single population and one spatial dimension. For the case
when $a$ tends to $0$ together with $\tau$ in such a way that the
ratio $a^3/\tau$ tends to finite limit $b$ it is possible to obtain
an equation for the probability density function $p(x,t)$ as follows 
(for more discussion see Appendix C):
\begin{equation}\label{ed2}
\frac{\partial p}{\partial t} + \frac{\partial}{\partial x} \bigg[ X(x) p \bigg ]
=\frac{b}{2}\frac{\partial^2 p}{\partial x^2} .
\end{equation}
For the general case of $N$ populations the equation for the
probability density function becomes
\begin{eqnarray}\label{ed3}
\frac{\partial p}{\partial t} + \sum_{i=1}^N \frac{\partial}{\partial x_i} \bigg[ X_i(x_1,
x_2,\dots,x_N) p \bigg ]= 
\frac{1}{2}\sum_{i=1}^N \sum_{j=1}^N b_{ij} \frac{\partial^2 p}{\partial x_i \partial x_j} ,
\end{eqnarray}
where $b_{ij}$ are again coefficients that characterize the random impulses.
\par
Another kind of problem that can be solved by this approach is to calculate the
mathematical expectation of the exit time. Let us again first discuss
the case of one population and one spatial dimension. We have a phase point that is
inside the interval $[\epsilon_1,\epsilon_2]$ ($\epsilon_1 < \epsilon_2$) and the
system is under the influence of the same random perturbations as described above.
The exit time is the time for which the phase point that was inside the
above interval at $t=0$ will leave this interval through $\epsilon_1$ or through
$\epsilon_2$. If we denote as $F(x)$ the mathematical expectation for the
exit time then $F(x)$ is a solution of the equation \cite{pav},\cite{jeschke} 
\begin{equation}\label{ed4}
\frac{b}{2} \frac{d^2 F}{dx^2} + X(x)\frac{dF}{dx}+1 =0 ,
\end{equation}
with boundary conditions $F(\epsilon_1)=F(\epsilon_2)=0$. For the case of many populations
the zero boundary conditions are on the entire border of the multidimensional
phase space area that has to be exited and the equation for the
probability density function of the exit time is
\begin{equation}\label{ed5}
\frac{1}{2} \sum_{i=1}^N \sum_{j=1}^N b_{ij} \frac{\partial^2 F}{\partial x_i 
\partial x_j}+ \sum_{i=1}^N X(x_1,x_2,\dots,x_N) \frac{\partial F}{\partial x_i} +1 =0 . 
\end{equation}

\par
Let us now apply this theory to Eq.(\ref{eq9}). We shall be interested in the
stationary distributions $p(\overline{\rho})$ connected to Eq.(\ref{eq9}),i.e.,
we shall be interested in the case when after a long
time the probability density function $p$ becomes stationary and depends
only on the spatial coordinate $\overline{\rho}$. This stationary case is important 
because of Observation 2 from Appendix C which states that each time dependent 
solution $p(x,t)$ of the Fokker-Planck equation (\ref{b7}) converges at $t \to \infty$ 
to the stationary distribution $p_0(x)$ from Eq.(\ref{b12}). In our case $x = \overline{\rho}$
and  $X(x) = \sum_{n_1=0}^L \alpha_{n_1} \overline{\rho}^{n_1}$. We shall
discuss two cases: (i) the case of single population where $\overline{\rho} \in [0, \infty)$;
and (ii) another case not connected directly to the population dynamics where
$\overline{\rho} \in (-\infty, \infty)$.\\
\emph{Case 1: $\overline{\rho} \in [0, \infty)$} \\
This case is connected directly to the population dynamics as the population density
can't be negative. in this case we have 
\newtheorem*{prop5}{Proposition 5}
\begin{prop5}
Let us discuss a system described by the equation
\begin{equation}\label{p1}
\frac{d \overline{\rho}}{dt}  = \sum_{n_1=0}^L \alpha_{n_1} \overline{\rho}^{n1}
\end{equation}
which is under the action of random impulses that occur after every
interval of time $\tau$ and each impulse causes the phase point of
the dynamical system described by Eqs. (\ref{p1}) to jump
through a distance $a$ along a random direction. Let
when $a$ tend to $0$ together with $\tau$ in such a way that the
ratio $a^3/\tau$ tends to finite limit $b$. Let in addition the following
requirements be fulfilled
\begin{eqnarray}\label{p1aa}
p(0) = 0; \ \rho \in [0, \infty)
\end{eqnarray}
Then the stationary p.d.f. $p(\overline{\rho})$ is
\begin{eqnarray}\label{p2x}
p(\overline{\rho}) = C \exp\left[\frac{2}{b} \sum_{n_1=0}^L
\alpha_{n_1}\frac{\overline{\rho}^{n_1+1}}{n_1+1} \right] \left \{ 1 - 
\int  {d \overline{\rho}} \ \exp \left [ - \frac{2}{b} \sum_{n_1=0}^L
\alpha_{n_1}\frac{\overline{\rho}^{n_1+1}}{n_1+1} \right ] \right \},
\end{eqnarray}
where the constant of integration $C$ is determined by the
normalization condition
\begin{equation}\label{p3}
\int_{-\infty}^{\infty} d \overline{\rho} \ p(\overline{\rho}) = 1 .
\end{equation}
\end{prop5}
\begin{proof}
The equation of the stationary distribution $p(\overline{\rho})$ is 
particular case of Eq. (\ref{ed2}). One integration of the equation for
the stationary distribution leads to
\begin{equation}\label{p3a}
\frac{b}{2} \frac{dp}{d\overline{\rho}} - \left[  p(\overline{\rho})\sum_{n_1=0}^L \alpha_{n_1} \overline{\rho}^{n1}
\right ] + C_1 =0
\end{equation}
Eq. (\ref{p3a}) can be easily integrated and the result is
\begin{eqnarray}\label{p3b}
p(\overline{\rho}) = \exp\left[\frac{2}{b} \sum_{n_1=0}^L
\alpha_{n_1}\frac{\overline{\rho}^{n_1+1}}{n_1+1} \right] \left \{ C - \frac{2 C_1}{b} 
\int  {d \overline{\rho}} \ \exp \left [ - \frac{2}{b} \sum_{n_1=0}^L
\alpha_{n_1}\frac{\overline{\rho}^{n_1+1}}{n_1+1} \right ] \right \},
\end{eqnarray}
where $C$ is a constant of integration. The condition $p(0)=0$ leads to
$C_1 = bC/2$ and distribution described by Eq. (\ref{p3b}) is reduced to
the distribution from Eq. (\ref{p2x}). 
\end{proof}

\emph{Case 2: $\overline{\rho} \in (-\infty, \infty)$} \\
\newtheorem*{prop6}{Proposition 6}
\begin{prop6}
Let us discuss a system described by the equation
\begin{equation}\label{p1}
\frac{d \overline{\rho}}{dt}  = \sum_{n_1=0}^L \alpha_{n_1} \overline{\rho}^{n1}
\end{equation}
which is under the action of random impulses that occur after every
interval of time $\tau$ and each impulse causes the phase point of
the dynamical system described by Eqs. (\ref{p1}) to jump
through a distance $a$ along a random direction. Let
when $a$ tend to $0$ together with $\tau$ in such a way that the
ratio $a^3/\tau$ tends to finite limit $b$. 
\begin{description}
\item[$\star$] \emph{Subcase 1:} \\
Let in addition the following requirements be fulfilled
\begin{eqnarray}\label{p1aa}
p(0) = A; \ \overline{\rho} \in (-\infty, \infty)
\end{eqnarray}
Then the stationary p.d.f. $p(\overline{\rho})$ is
\begin{eqnarray}\label{p2y}
p(\overline{\rho}) =  \exp\left[\frac{2}{b} \sum_{n_1=0}^L
\alpha_{n_1}\frac{\overline{\rho}^{n_1+1}}{n_1+1} \right] \left \{ C - (C-A) 
\int  d \overline{\rho} \ \exp \left [ - \frac{2}{b} \sum_{n_1=0}^L
\alpha_{n_1}\frac{\overline{\rho}^{n_1+1}}{n_1+1} \right ] \right \},
\end{eqnarray}
where the constant of integration $C$ is determined by the
normalization condition
\begin{equation}\label{p3}
\int_{-\infty}^{\infty} d \overline{\rho} \ p(\overline{\rho}) = 1 .
\end{equation}
\item[$\star \star$] \emph{Subcase 2:} \\
Let in addition the following
requirements be fulfilled
\begin{eqnarray}\label{p1a}
\frac{1}{p(0)}\frac{dp}{d \overline{\rho}}\mid_{\overline{\rho}=0} = \frac{2 \alpha_0}{b};
\ \overline{\rho} \in (-\infty, \infty)
\end{eqnarray}
Then the stationary p.d.f. $p(\overline{\rho})$ is
\begin{eqnarray}\label{p2}
p(\overline{\rho}) = C \exp\left[\frac{2}{b} \sum_{n_1=0}^L
\alpha_{n_1}\frac{\overline{\rho}^{n_1+1}}{n_1+1} \right] ,
\end{eqnarray}
where the constant of integration $C$ is determined by the
normalization condition
\begin{equation}\label{p3}
\int_{-\infty}^{\infty} d \overline{\rho} \ p(\overline{\rho}) = 1 .
\end{equation}
\end{description}
\end{prop6}
\begin{proof}
\emph{Subcase 1} \\
The equation of the stationary distribution $p(\overline{\rho})$ is 
particular case of Eq. (\ref{ed2}). One integration of the equation for
the stationary distribution leads to Eq. (\ref{p3a}).
Eq. (\ref{p3a}) can be easily integrated and the result is Eq. (\ref{p3b}).
where $C$ is a constant of integration. The condition $p(0)=A$ leads to
$C_1 = b(C-A)/2$ and distribution described by Eq. (\ref{p3b}) is reduced to
the distribution from Eq. (\ref{p2y}).\\ 
\emph{Subcase 2} \\
The integration of Eq.(\ref{ed2}) leads to
Eq.(\ref{p3a}).
where $C_1$ is a constant of integration. This constant is equal to $0$ because
of the condition (\ref{p1a}).
With $C_1=0$ we can continue the integration of Eq.(\ref{p3a}) and
the result is (\ref{p2}) where the constant of integration $C$ is determined by the
normalization condition (\ref{p3}).
\end{proof}
We note that Eq. (\ref{p1a}) in combination with Eq. (\ref{p2}) mean that  $\sum_{n_1=0}^L \alpha_{n_1}=0$.
In addition $f(\overline{\rho})$ must tend to $0$ when $\overline{\rho} \to \pm \infty$. The dominant term
at large values of $\overline{\rho}$ is $\alpha_L \overline{\rho}^L$. Then $\alpha_L$ must be negative
(to ensure $f \to 0$ at large positive values of $\overline{\rho}$ and $L$ must be odd (to ensure 
$f \to 0$ at large negative values of $\overline{\rho}$).
\par
Several examples  for statistical distributions $f(\overline{\rho})$ are shown in Fig.2.
\begin{figure}[t]
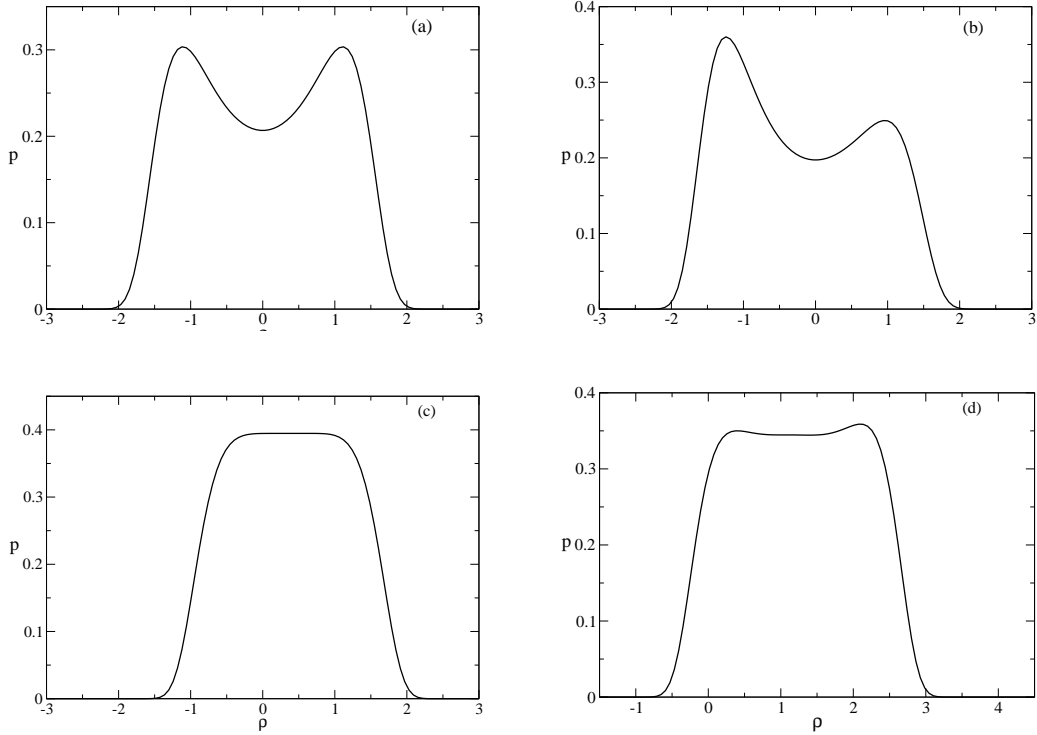

\centering
\includegraphics[width=2.5in]{distrib_a.eps}\hskip1cm
\includegraphics[width=2.5in]{distrib_b.eps}\vskip0.5cm
\includegraphics[width=2.5in]{distrib_c.eps}\hskip1cm
\includegraphics[width=2.5in]{distrib_d.eps}\\
\caption{Several profiles of $p(\overline{\rho})$ from Eq.(\ref{p2}) (in the figures
$\overline{\rho}$ is denoted as $\rho$). Figure (a):
$b=2$; $\alpha_0 =0$; $\alpha_1 = 1$; $\alpha_2 = -0.4$; $\alpha_3 =0.2$; 
$\alpha_4 = 0$; $\alpha_5 =-0.5$. Figure (b):
$\alpha_0 =0$; $\alpha_1 =1$; $\alpha_2 = 0$; $\alpha_3 =0.2$; $\alpha_4 = 0$; 
$\alpha_5 =-0.5$; $b=2$. Here the fixed points are 3.
The two maxima of the p.d.f. distribution are centered around the two stable fixed points and
the minimum is centered on the unstable fixed point $\overline{\rho}=0$. Figure (c):
$\alpha_0=0.003564$; $\alpha_1 =-0.006084$; $\alpha_2 = 0.3975$;$\alpha_3=-1.234$; $\alpha_4 =1.81$; 
$\alpha_5 = -1$; $b=2$. Figure (d): $\alpha_0=1.2936$; $\alpha_1 =-7.2436$; $\alpha_2 = 14.58$; $\alpha_3 =-13.63$; 
$\alpha_4 =6$; $\alpha_5 = -1$; $b=2$. The probability distribution has 3 maxima (at $\overline{\rho}=0.4,1.1,2.1$) 
and two minima.}
\label{fig1a}
\end{figure}
Let us note here the tunnel phenomenon which arises because of the presence of
fluctuations. The existence of the distributions $p(\rho,t)$ and $p_0(\rho)$ means that
in the course of the time each value of the density $\rho$ can be reached. Then if for
an example at the initial moment the system trajectory in the phase space is close to
a fixed point of the ODE without added fluctuations then in the course of the time
the phase point can leave this area and to travel to phase space area that is close
to another fixed point. This tunnel phenomena is closely connected to the role played
by fluctuations in the case when a bifurcation happens in the studied system.
\par
Let us now calculate the exit time expectation on the basis of Eq.(\ref{ed4}). 
\newtheorem*{prop7}{Proposition 7}
\begin{prop7}
let us discuss the system described by Eq.(\ref{p1}). The distribution $F_q(\overline{\rho})$
from the initial position $\overline{\rho}$ to the position $q < \overline{\rho}$ is
\begin{eqnarray}\label{ed10}
F_q(\overline{\rho}) = \int_{q}^{\overline{\rho}} d \xi \ \exp \bigg( 
- \frac{2}{b} \sum_{n_1=0}^L \alpha_{n_1} \frac{\xi^{n_1+1}}{n_1+1}\bigg)
\bigg[\frac{2}{b} \int_{\xi}^\infty d \eta \ \exp \bigg(\frac{2}{b}
\sum_{n_1=0}^L \alpha_{n_1} \frac{\eta^{n_1+1}}{n_1+1} \bigg) \bigg] .
\end{eqnarray}
when
\begin{description}
\item[($^\star$)]
$F(\overline{\rho}=q) = 0$ ,
\item[($^{\star \star}$)]
$F(\overline{\rho},q)$ increases in the slowest possible manner as
$\overline{\rho} \to \infty$ .
\end{description}
\end{prop7}
\begin{proof}
We calculate the distribution for exit time from the initial
position $\overline{\rho}$ to a position $q < \overline{\rho}$.
In the discussed case again  $X(x) = \sum_{n_1=0}^L \alpha_{n_1} \overline{\rho}^{n_1}$.
One integration of  Eq.(\ref{ed4}) leads to the equation
\begin{eqnarray}\label{ed9}
\frac{d F}{d \overline{\rho}}= \exp(-\psi(\overline{\rho})) \bigg(
C_1 + \int_{\overline{\rho}}^\infty d \xi \ \frac{2}{b} \exp(\psi(\xi)) \bigg);\ 
\psi (\overline{\rho})=\frac{2}{b}\int_{\overline{\rho}}^\infty d \xi X(\xi)
\end{eqnarray}
The relationship ($^{\star \star}$) requires $C_1 = 0$ (as the corresponding term in Eq. (\ref{ed9}) vanishes
and the growth of $\frac{d F}{d \overline{\rho}}$ is as slow as possible),  and the integration of
Eq.(\ref{ed9}) leads to the result (\ref{ed10})
\end{proof}
\begin{figure}[t]
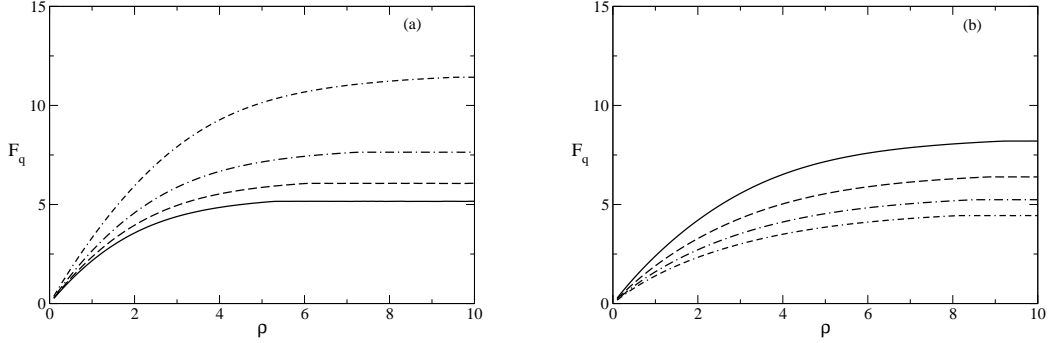

\centering
\includegraphics[width=2.5in]{exit1.eps} \hskip1cm
\includegraphics[width=2.5in]{exit2.eps}
\caption{Exit time expectations for $q=0$ (which means extinction of the population)
calculated on the basis of Eq.(\ref{ed10}). $\rho$ on the horizontal axis is equal to
$\overline{\rho}$ from Eq.(\ref{ed10}). For all curves $b=2$. Figure (a): influence of the
value of $\alpha_3$ on the exit time.  $\alpha_0 = 0.01$; $\alpha_1=0.2$; $\alpha_2=0.1$. 
Solid line: $\alpha_3=-0.05$; dashed line: $\alpha_3=-0.04$; dot-dashed line: $\alpha_3 = -0.03$;
dot-double dashed line: $\alpha_3=-0.02$. Figure (b): influence of the value of $\alpha_1$ on th exit time.
$\alpha_0=0.01$;$\alpha_2=0.1$; $\alpha-3=-0.02$. Solid line: $\alpha_1=-0.3$; dashed line: $\alpha_1=-0.4$;
dot-dashed line: $\alpha_1=-0.5$; dot-double dashed line: $\alpha_1=-0.6$. }
\label{fig2}
\end{figure}
Fig. 3 shows the dependence of the exit time expectation on the
population density and coefficients of the model equation for the
case $L=3$. The negative values of $\alpha_{1,3}$ make extinction be expected sooner whereas
the positive values of the other two parameters can delay the extinction.
The theory can be easily applied for the case of
system of many interacting populations but even in the simplest
one-dimensional case the integral from Eq.(\ref{ed10}) must be calculated
numerically.
\section{Conclusion}
In this paper we have discussed two aspects of population dynamics. First we
have presented a model of the space-time dynamics of the system of interacting
population in 2 spatial dimensions. For the most simple case of one spatial 
dimension and for one population we have obtained exact traveling-wave solution of
the model nonlinear PDE by means of the recently developed modified method of
simplest equation for obtaining exact and approximate solutions of nonlinear PDEs.
The obtained exact solution describes the propagation of changes of the population
density in the space. The generalization of this theory to the case of many populations
is straightforward and describes the spreading of coupled waves of changes of densities
of the studied populations. The case of three populations is discussed in the
paper. The second discussed
aspect of the population dynamics was connected to the influence of the random fluctuations
on the population densities.  The presence of fluctuations leads to description in terms of
probability density functions for the population densities. The discussed general theory
is illustrated again for the simplest possible case of one population in two aspects:
calculation of probability density functions and calculation of the expected extinction
time. As expected from the theory of diffusion Markov processes the minima and 
maxima of the obtained probability density functions are exactly
at the fixed points of the corresponding non-perturbed model system of differential
equations. The expected extinction time strongly depends on the coefficients of the
model equations. Finally several results are obtained that are not connected to the
theory of interacting populations but may be interesting for other cases modeled by
nonlinear PDEs with polynomial nonlinearities. 
\section*{Acknowledgment}
This research was partially supported by the Fund of Scientific Researches of
Republic of Bulgaria under contract DO 02-338/22.12.2008 in the scope of which
the averaging applied in section II B has been developed and used.

\begin{appendix}
\section{Modified method of simplest equation}
\begin{figure}[t]\label{fig3}
\begin{center}
\includegraphics[scale=0.4]{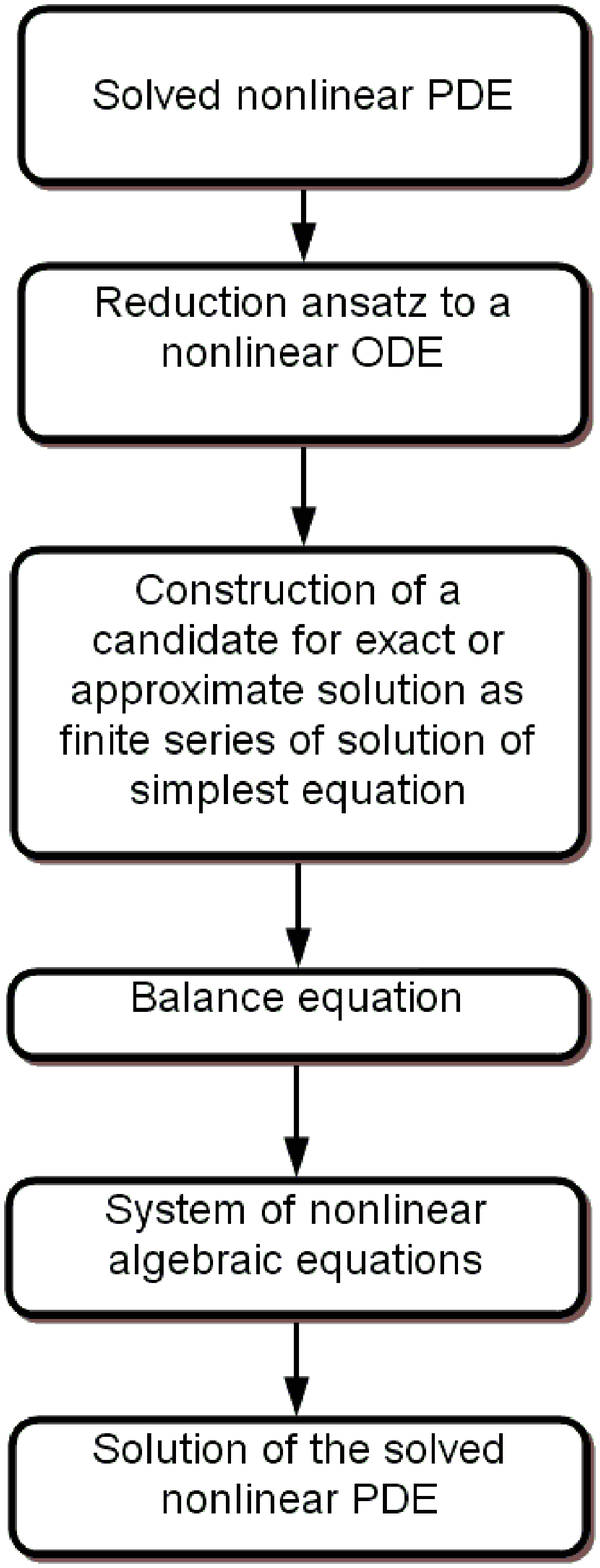}
\end{center}
\caption{The modified method of simplest equation }
\end{figure}
There are many approaches for obtaining exact analytic solutions 
of nonlinear partial differential equations \cite{her}-\cite{vm96}. In this paper
we use the modified method of simplest equation.
The schema of application of the modified method of simplest equation
is shown in Fig.4. The method of simplest equation \cite{kudr05x}-\cite{kudr08} 
is based on the fact that after application
of appropriate ansatz large class of  NPDEs can be reduced to ODEs of the kind
\begin{equation}\label{et2}
{\cal{P}}\left( F(\xi),\frac{d F}{d \xi}, \frac{d^2F}{d \xi^2},\dots \right)=0,
\end{equation}
and for some equations of the kind (\ref{et2}) particular solutions can be obtained
which are finite series
\begin{equation}\label{et3}
F(\xi) = \sum_{i=0}^P a_i [\Phi(\xi)]^i,
\end{equation}
constructed by solution $\Phi(\xi)$ of more simple equation referred to as
simplest equation. The simplest equation can be the equation of Bernoulli, equation
of Riccati, etc. The substitution of Eq. (\ref{et3}) in Eq. (\ref{et2})
leads to the polynomial equation
\begin{equation}\label{et4}
{\cal P}= \sigma_0 + \sigma_1 \Phi + \sigma_2 \Phi^2 + \dots + \sigma_r \Phi^r =0 ,
\end{equation}
where the coefficients $\sigma_i$, $i=0,1,\dots,r$ depend on the parameters of the
equation and on the parameters of the solutions. Equating all these coefficients to $0$,
i.e.,  by setting
\begin{equation}\label{et5}
\sigma_i =0, i=1,2,\dots,r ,
\end{equation}
one obtains a system of nonlinear algebraic equations. Each solution of this system
leads to a solution of kind (\ref{et3}) of Eq. (\ref{et2}).
\par
In order to ensure non-trivial solution by the above method we have to ensure that
$\sigma_r$ contains at least two terms. To do this we have to balance the highest
powers of $\phi$ that are obtained from the different terms of the solved equation of
kind (\ref{et2}). As a result of this we obtain an additional equation between some of the
parameters of the equation and the solution. This equation is called balance equation \cite{vdk10}-\cite{vit11a}.
\section{Maple program for solving the coupled waves case}
Here we present a Maple program for obtaining exact solution of a system of
equations for description of coupled waves in a system of 3 populations discussed in Sec....
The program has the following parts.\\

{\bf Part 1: The equations}\\
1.$>eq1:=-(v+D11)*diff(rho1(xi),xi)-D12*diff(rho2(xi),xi)-D13*diff(rho3(xi),xi)-$\\
 $r10*rho1(xi)+r10*alpha110*rho1(xi)^2-r10*alpha120*rho1(xi)*rho2(xi)-$\\
 $r10*alpha130*rho1(xi)*rho3(xi);$
2.$>eq2:=-D21*diff(rho_1(xi),xi)-(v+D22)*diff(rho2(xi),xi)-D23*diff(rho_3(xi),xi)-$\\
 $r20*rho2(xi)+r20*alpha210*rho1(xi)*rho2(xi)-r20*alpha220*rho_2(xi)^2-$\\
 $r20*alpha230*rho2(xi)*rho3(xi);$\\
3.$>eq3:=-D31*diff(rho1(xi),xi)-D32*diff(rho2(xi),xi)-(v+D33)*diff(rho3(xi),xi)-$\\
 $r30*rho3(xi)+r30*alpha310*rho1(xi)*rho3(xi)-r30*alpha320*rho2(xi)*rho3(xi)-$\\
 $r30*alpha330*rho3(xi)^2;$\\

{\bf Part 2: Substitution of the relationships for $\rho_{1,2,3}$ in the equations} \\
4.$>rho1(xi):=a0+a1*Phi(xi);$\\
5.$> rho2(xi):=b0+b1*Phi(xi);$\\
6.$>rho3(xi):=c0+c1*Phi(xi);$\\
7.$>eq1;eq2;eq3;$\\

{\bf Part 3: Substitution of the relationship for $\frac{d \Phi}{d \xi}$ in the equations} \\
8.$>eq1a:=subs(diff(Phi(xi),xi)=a*Phi(xi)^2+b*Phi(xi)+c,eq1);$\\
9.$>eq2a:=subs(diff(Phi(xi),xi)=a*Phi(xi)^2+b*Phi(xi)+c,eq2);$\\
10.$>eq3a:=subs(diff(Phi(xi),xi)=a*Phi(xi)^2+b*Phi(xi)+c,eq3);$\\

{\bf Part 4: Extracting the system of nonlinear algebraic equations}\\
11.$>eq1b:=collect(eq1a,Phi(xi));$\\
12.$>eq2b:=collect(eq2a,Phi(xi));$\\
13.$>eq3b:=collect(eq3a,Phi(xi));$\\
14.$>e1:=coeff(eq1b,Phi(xi)^2);$\\
15.$>e2:=coeff(eq1b,Phi(xi));$\\
16.$>e3:=-(v+D11)*a1*c-r10*alpha120*a0*b0-D13*c1*c-D12*b1*c-$\\
   $r10*alpha130*a0*c0-r10*a0+r10*alpha110*a0^2;$\\
17.$>e4:=coeff(eq2b,Phi(xi)^2);$\\
18.$>e5:=coeff(eq2b,Phi(xi));$\\
19.$>e6:=-D21*a1*c-r20*alpha220*b0^2-D23*c1*c-(v+D22)*b1*c-$\\
   $r20*alpha230*b0*c0-r20*b0+r20*alpha210*a0*b0;$\\
20.$>e7:=coeff(eq3b,Phi(xi)^2);$\\
21.$>e8:=coeff(eq3b,Phi(xi));$\\
22.$>e9:=-D31*a1*c-r30*alpha320*b0*c0-(v+D33)*c1*c-D32*b1*c-$\\
   $r30*alpha330*c0^2-r30*c0+r30*alpha310*a0*c0;$\\

{\bf Part 5: Simplifying assumptions}\\
23.$>r10:=1;r20:=1;r30:=1;alpha110:=1;alpha220:=1;alpha330:=1;$\\
   $alpha120:=1;alpha130:=1;alpha210:=1;alpha230:=1;alpha310:=1;$\\
   $alpha320:=1;D21:=D12;D31:=D13;D32:=D23;D22:=D11;D33:=D11;$\\
   $ D12:=1;D13:=1;D23:=1;$\\
24.$>e1;e2;e3;e4;e5;e6;e7;e8;e9;$\\

{\bf Part 6: Solution of the system of nonlinear algebraic equations}\\
25.$>sol1:=solve(e1,c_1);$\\
26.$>c1:=sol1;$\\
27.$>sol2:=solve(e2,v);$\\
28.$>v:=sol2;$\\
29.$>sol3:=solve(e3,c);$\\
30.$>c:=sol3;$\\
31.$>sol4:=solve(e4,a);$\\
32.$>a:=sol4[1];$\\
33.$>sol5:=solve(e5,a_0);$\\
34.$>a_0:=sol5;$\\
35.$>sol6:=solve(e8,b_0);$\\
36.$>b0:=sol6;$

By this program one obtains relationships for the parameters $a_0$,$b_0$,$c_1$,$a$,$c$,$v$.

\section{Fluctuations, diffusion Markov process and Fokker-Planck equation}
In the paper we discuss equations of the kind
\begin{equation}\label{b1}
\frac{dx}{dt} = X[x(t)]+ \eta(t)
\end{equation}
where the process $\eta(t)$ models the small fluctuations. Let $\eta(t)=\sigma \xi(t)$ where
$\sigma >0$ is the intensity factor and the covariance function of the process $\xi(t)$ be a
$\delta$-function $E[\xi(t) \xi'(t)]= \delta(t-t')$. If in addition the expected value of $\xi(t)$
is $0$: $E[\xi(t)]=0$ the process $\xi(t)$ is called white noise and the equation
\begin{equation}\label{b2}
\frac{dx}{dt} = X[x(t)]+ \sigma \xi(t); \ \ x(0) = x_0
\end{equation}
is called Langevin equation.
\par
$\xi(t)$ can be written as time derivative of a Wiener process $W_t$ (for this process $W_0=0$;
the function $W_t(t)$ is almost surely continuous; and $W_t$ has independent increments
$W_t-W_s$ $(0 \le s <t)$ which are normally distributed with expected value $0$ and variance 
equal to $t-s$):
\begin{equation}\label{b3}
\xi(t)=\frac{d W_t}{dt} \ \to W_t = \int_0 ds \ \xi(s)
\end{equation}
 Then the Langevin equation can be written in the form
\begin{equation}\label{b4}
dx_t = X(x_t)dt + \sigma dW_t; \ \ x_0:{\rm random}
\end{equation}
After one integration of the Langevin equation one obtains
\begin{equation}\label{b5}
x_t(\omega) = x_0(\omega) + \int_0 ds X[x_s(\omega)] + \sigma W_t(\omega)
\end{equation}
as the white noise is $\delta$-correlated the solution of Eq.(\ref{b5}) is a homogeneous Markov
process. The infinitesimal generator $A$ of the solution process $x_t$ 
\cite{jeschke}, \cite{wentzell} is a differential operator of second order
\begin{equation}\label{b6}
(Ag)(x)=f(x)g'(x) + \frac{\sigma^2}{2}g''(x)
\end{equation}
The form of the infinitesimal operator $A$ is important as it is determines the
form of the Fokker-Planck equation for the one dimensional distribution $p(x,t)$ connected to the
solution process (\ref{b5}). This Fokker-Planck equation is 
\begin{equation}\label{b7}
\frac{\partial p(x,t)}{\partial t} = - \frac{\partial}{\partial x} \left[ X(x) p(x,t) \right] +
\frac{\sigma^2}{2} \frac{\partial^2 p(x,t)}{\partial x^2}; \ \ p(x,0) = p_0(x) 
\end{equation}
We note that if we set $\sigma^2 = b$ we obtain Eq.(\ref{ed2}) from the main text of the paper.
By setting
\begin{equation}\label{b8}
J_p = p X - \frac{\sigma^2}{2} \frac{\partial p}{\partial x}
\end{equation}
we can write Eq.(\ref{b7}) in the form
\begin{equation}\label{b9}
\frac{\partial p}{\partial t} + \frac{\partial J_p}{\partial x} = 0
\end{equation}
Eq.(\ref{b9}) is a balance equation for the probability. There is no production
of probability and the balance equation shows that the transport of probability
along $x$ happens by 'flow', determined by the 'velocity' field $X$ and 'diffusion'
determined by $ - \cfrac{\sigma^2}{2} \cfrac{\partial p}{\partial x}$. This 'diffusion'
tries to decreases the differences in the probability distribution. 
\par
Because of the above-mentioned the considered class of Markov processes are called
diffusion processes. $X$ is the drift of the diffusion process and $\sigma^2$ is
called diffusion coefficient of the diffusion process. 
\par
We finish this short discussion by two observations that are important for the text
in the body of the paper \cite{jeschke}
\begin{flushleft}
{\bf Observation 1}: 
\end{flushleft}
{\it In the stationary case $p(x,t) \to p(x,0) = p_0(x)$ the stochastic
system described by Eq.(\ref{b4}) spends much time around the stable fixed points of the
corresponding deterministic system
$$
\frac{dx}{dt} = X(x)
$$
}
Observation 1 is a consequence of the fact that for the stationary case the Fokker-Planck
equation (\ref{b7}) becomes
\begin{equation}\label{b10}
0 = - \frac{d}{dx}[X(x) p_0(x)] + \frac{\sigma^2}{2} \frac{d^2 p_0(x)}{d x^2}
\end{equation}
After one integration one obtains
\begin{equation}\label{b11}
J_p^0 = - X(x) p_0(x) + \frac{\sigma^2}{2} \frac{d p_0(x)}{dx} = {\rm const}
\end{equation}
The integration constant is equal to $0$ because of the boundary condition $J_p^0 = 0 \mid_{\pm \infty}$
and for $p_0(x)$ one obtains
\begin{equation}\label{b12}
p_0(x) = \cfrac{\exp\left[- \cfrac{2}{\sigma^2} V(x) \right]}{\int_{-\infty}^{\infty} dx \ 
\exp\left[- \cfrac{2}{\sigma^2} V(x) \right]}
\end{equation}
where 
\begin{equation}\label{b13}
V(x) = - \int^x dz \ X(z)
\end{equation}
\par
The local extrema of $p_0(x)$ are given by
\begin{equation}\label{b14}
\frac{dp_0}{dx}\mid_{x_0}=0 \xrightarrow{(\ref{b12})}\frac{dV}{dx}\mid_{x_0}=0 \xrightarrow{(\ref{b13})}
f(x_0) = 0
\end{equation}
Thus the extrema of $p_0(x)$ coincide to the fixed points of the corresponding deterministic
differential equation. The maxima of $p_0(x)$ are located at the stable fixed points and the
minima of $p_0(x)$ are located at the unstable fixed points. 
\par
The second observation is connected to the importance of the stationary distribution $p_0(x)$:
\begin{flushleft}
{\bf Observation 2}:
\end{flushleft}
{\it
Each time dependent solution $p(x,t)$ of the Fokker-Planck equation (\ref{b7}) converges at $t \to \infty$ 
to the stationary distribution $p_0(x)$ from Eq.(\ref{b12}) (if $p_0(x)$ exists).
}
\par
In order to show that Observation 2 is true we shall use the technique of the Lyapunov
functional $H(t)$. Let us discuss the functional
\begin{equation}\label{b15}
H(t) = \int_{-\infty}^{\infty} dx \ p(x,t) \ln \frac{p(x,t)}{p_0(x)}
\end{equation}
where $p(x,t)$ is arbitrary solution of the Fokker-Planck equation (\ref{b7}). 
Using the normalization $\int_{-\infty}^{\infty} dx \ p(x,t) = 1$ for each $t>0$ and the fact
that $\ln(1/y) \ge 1 - y$ fro $y>0$ one can write the inequality
\begin{equation}\label{b16}
H(t) = \int_{-\infty}^{\infty} dx \ p(x,t) \left [ \ln \frac{p(x,t)}{p_0(x)} + \frac{p_0(x)}{p(x,t)} -1 \right ]
\ge 0
\end{equation}
where the equality arises for $p(x,t) = p_0(x)$. We shall show that $dH/dt \le 0$ for each $t$. We take the
derivative of $H$ with respect to $t$ and by means of the Fokker-Plank equation (\ref{b7}) after some
calculations we obtain the relationships
\begin{eqnarray}\label{b17}
\frac{dH}{dt} = \int_{-\infty}^{\infty} dx \left[ - \frac{\partial}{\partial x}[X(x) p(x,t)]+ \frac{\sigma^2}{2}
\frac{\partial^2 p(x,t)}{\partial x^2} \right] \left [ \ln \frac{p(x,t)}{p_0(x)}\right ] = \nonumber \\
-\frac{\sigma^2}{2}
\int_{-\infty}^{\infty} dx p(x,t) \left[\frac{p_0(x)}{p(x,t)} \frac{\partial}{\partial x} \left(\frac{p_0(x)}{p(x,t)} 
\right)\right]^2
\end{eqnarray}
The last integral in (\ref{b17}) is positive for $p \ne p_0$ and it is equal to $0$ only when $p(x,t)=p_0(x)$.
Then $\cfrac{dH}{dt} \le 0$ and $\cfrac{dH}{dt}=0$ only when $p(x,t)=p_0(x)$. From (\ref{b16}) and from
$\cfrac{dH}{dt} \le 0$ it follows that
\begin{equation}\label{b18}
\lim_{t \to \infty} p(x,t) = p_0(x)
\end{equation}
which is exactly the essence of {\bf Observation 2}.

\end{appendix}





\bibliographystyle{elsarticle-num}
\bibliography{<your-bib-database>}



\end{document}